\documentclass[twocolumn,showpacs,aps,superscriptaddress]{revtex4-1}
\usepackage{amsmath}
\usepackage[switch, modulo]{lineno}
\usepackage{graphicx}
\usepackage{longtable}
\usepackage{ltablex}
\include{def}
\def\Journal#1#2#3#4{{#1} {#2} (#3) #4}

\def\JPG{J.Phys.G}

\def\EPJ{Eur.Phys.J.}

\def\NPA{Nucl.Phys.A}

\def\PLB{Phys.Lett.B}

\def\PRL{Phys.Rev.Lett.}
\def\PRC{Phys.Rev.C}
\def\PRD{Phys.Rev.D}

\def\ZPA{Z.Phys.A}

\def\EPJ{Eur.Phys.J.}

\def\CPC{Comp.Phys.Comm.}
\def\be{\begin{equation}}
\def\ee{\end{equation}}

\begin{document}

\title{On similarities as a function of system size\\
in heavy ion collisions}

\author{M.Petrovici}
\affiliation{National Institute for Physics and Nuclear Engineering - IFIN-HH\\
Hadron Physics Department\\
Bucharest - Romania}
\affiliation{Faculty of Physics, University of Bucharest}
\author{A.Lindner}
\affiliation{National Institute for Physics and Nuclear Engineering - IFIN-HH\\
Hadron Physics Department\\
Bucharest - Romania}
\affiliation{Faculty of Physics, University of Bucharest} 
\author{A.Pop} 
\affiliation{National Institute for Physics and Nuclear Engineering - IFIN-HH\\
Hadron Physics Department\\
Bucharest - Romania}
\date{\today}

\begin{abstract}
Qualitative and quantitative similarities as a function of the system size in heavy ion collisions from low energy dissipative collisions, collective expansion of compressed baryonic matter up to the geometrical scaling evidenced at the highest energies presently attainable at LHC, are presented. 
\end{abstract}
\maketitle
\section{Introduction}
Intensive theoretical and experimental effort during the last four decades in the field of heavy ion collisions at low, relativistic and ultra-relativistic energies, has shown that transient pieces of matter at high excitation energy, density and temperature can be produced. However, one has to consider that even in the case of colliding the heaviest nuclei the size of the created system is rather small, increasing the violence of the collision the initial state becomes highly non-homogenous and dynamical effects play a crucial role. Studies of the influence of the dynamics and system size effects on the extracted information become mandatory. 
Therefore, collision of symmetric or asymmetric systems of different mass 
were studied in all energy domains, starting from energies well below the Fermi values up to the LHC energies where the parton distribution of nucleons at low x and $Q^2$ values starts to play an essential role. In the present lecture we review some of the similarities as a function of the system size
experimentally evidenced in different energy domains. In section 2 is presented such a similarity in terms of the width of the charge distribution as a function of total kinetic energy loss (TKEL) divided by the grazing angular momentum ($l_{gr}$) in dissipative heavy ion collisions at different collision energies, below 
15 A$\cdot$MeV and for different masses of the colliding nuclei. Similarities in terms of collective expansion based on a hybrid phenomenological model for compressed and hot baryonic matter produced in the collision of symmetric systems of different mass at collision energies above 
100 A$\cdot$MeV are summerised in Section 3. Section 4 is dedicated to similarities in terms of geometrical scaling for different masses of nuclei in symmetric collisions at RHIC and LHC energies. Conclusions are presented in Section 5.       

\section{Charge variance systematics in heavy ion dissipative collisions}

  The study of  dissipative collisions was one of the main topics at the beginning of the heavy-ion research field \cite{Bock}. Among many interesting evidenced phenomena,
it was clearly shown that at projectile energies below $\sim$15 A$\cdot$MeV,
via such processes, where the formation of a di-nuclear system and the stochastic 
nucleon exchange play the major role, the energy transfer from the relative motion to the internal excitation of the colliding partners can reach values at which the de-excitation time of statistical emission of different species starts to be shorter than the lifetime of the di-nuclear system 
\cite{Stef} and the charge/mass distribution in the two-body exit channel covers the full range of the mass asymmetry \cite{Gral}.     
 Based on the experimental results obtained in heavy ion collisions at projectile energies below 8 A$\cdot$MeV, it was evidenced a correlation between the width of charge distribution as a function of the total final kinetic energy divided by the grazing angular momentum \cite{Wol}. All measured systems at different collision energies show the same slope in such a representation, the offsets being related to the ratio of the collision energy versus entrance channel angular momentum, this dependence being also nicely correlated. Systematic studies of dissipative collisions at a higher collision energy \cite{Pet1}, $\sim$15 A$\cdot$MeV, from asymmetric to symmetric combinations, confirmed this scaling, the slope of the logarithm of the charge distribution width as a function of the total final kinetic energy divided by the grazing angular momentum being the same within the error bars. To what extent such a scaling also works for dissipative collisions of light nuclei became an interesting question. 
\begin{figure*}[t!]
\includegraphics[width=0.80\linewidth]{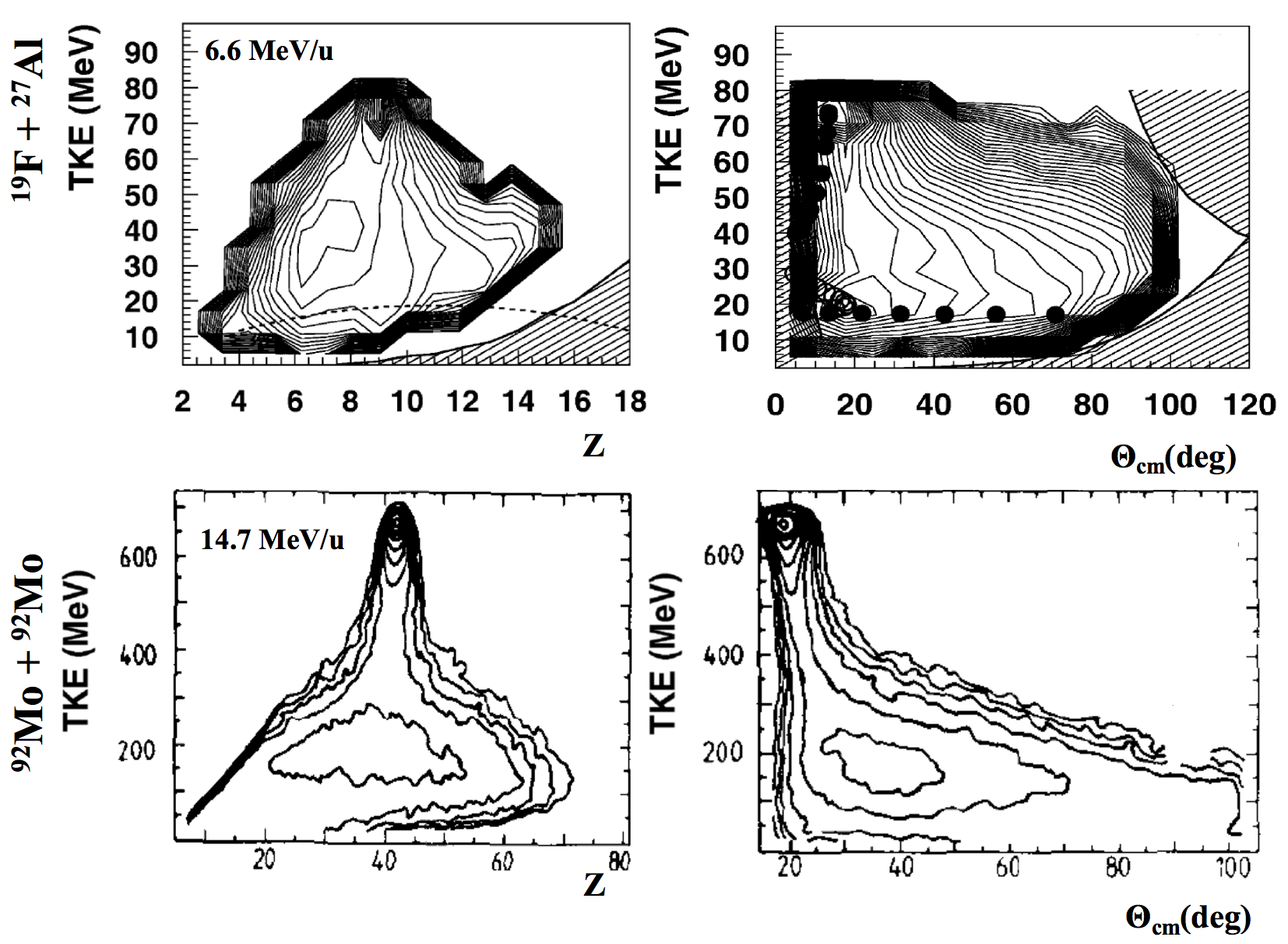}
\caption{Left: Diffusion plots; Right: TKE-$\Theta_{cm}$ correlation plots for $^{19}F$+$^{27}Al$ at 6.6 A$\cdot$MeV \cite{Pet2,Pop1,Pop2} (top row) and  $^{92}Mo$+$^{92}Mo$ at 14.7 A$\cdot$MeV \cite{Pet1} (bottom row) dissipative collisions.}
\end{figure*}
A series of experiments done using light nuclei at
projectile energies below 7 A$\cdot$MeV 
\cite{Pet2, Pop1,Pop2} revealed clear qualitative similarities in many respects with the trends 
observed in medium and heavy mass ion collisions. As an example, in Fig.1 are presented diffusion and total kinetic energy (TKE) versus center-of-mass scattering angle ($\Theta_{cm}$) correlation plots for 
$^{19}F$+$^{27}Al$ at 6.6 A$\cdot$MeV and $^{92}Mo$+$^{92}Mo$ at 14.7 A$\cdot$MeV dissipative collisions. For both systems, quite different in mass and collision energy, one can observe a
qualitative similarity in the trends of charge and polar angle in the center of mass of the two-body exit channel distributions as a function of final total kinetic energy. In Fig.2a, by representing the logarithm of  $\sigma_Z^2$, obtained by fitting the charge distribution of the final two-body channels products for a given value of 
TKEL=$TKE^{in}$-$TKE^{out}$ with a Gaussian function, as a function of $\frac{TKEL}{l_{gr}}$,  a linear dependence, similar with the results obtained for much heavier masses and larger collision energies, is shown. The linear fit gives a slope of 5.93$\pm$0.18 $\hbar$/MeV, to be compared with 6.07 $\hbar$/MeV 
from \cite{Wol} and 5.25 $\hbar$/MeV from \cite{Pet1}. It was shown in \cite{Pop1} that introducing an offset which is dependent on the
incident channel characteristics, i.e.:
\begin{equation}
F\sim\frac{TKE^{in}-V_{coul}^{in}}{\mu\cdot l_{gr}}(A/Z)^2
\end{equation}
where $TKE^{in}-V_{coul}^{in}$ is the total kinetic energy in the ingoing channel above the Coulomb energy and $\mu$ is the reduced mass of the colliding system, $ln(F\cdot\sigma_Z^2)$ for 20 different colliding systems and energies \cite{Pet1,Pet2,Pop1,Pop2,Schr,Dye,Wol1,Hoo,Birk} follow the same trend as a function of $TKEL/l_{gr}$. The result can be followed in Fig.2b. This suggests that the underlying mechanism of dissipative collisions is independent on the mass values, mass asymmetry and collision energy within the energy range where fast processes such as preequilibrium emission, more than two body processes and multifragmentation play a minor role.    
\begin{figure*}[ht]
\includegraphics[width=0.80\linewidth]{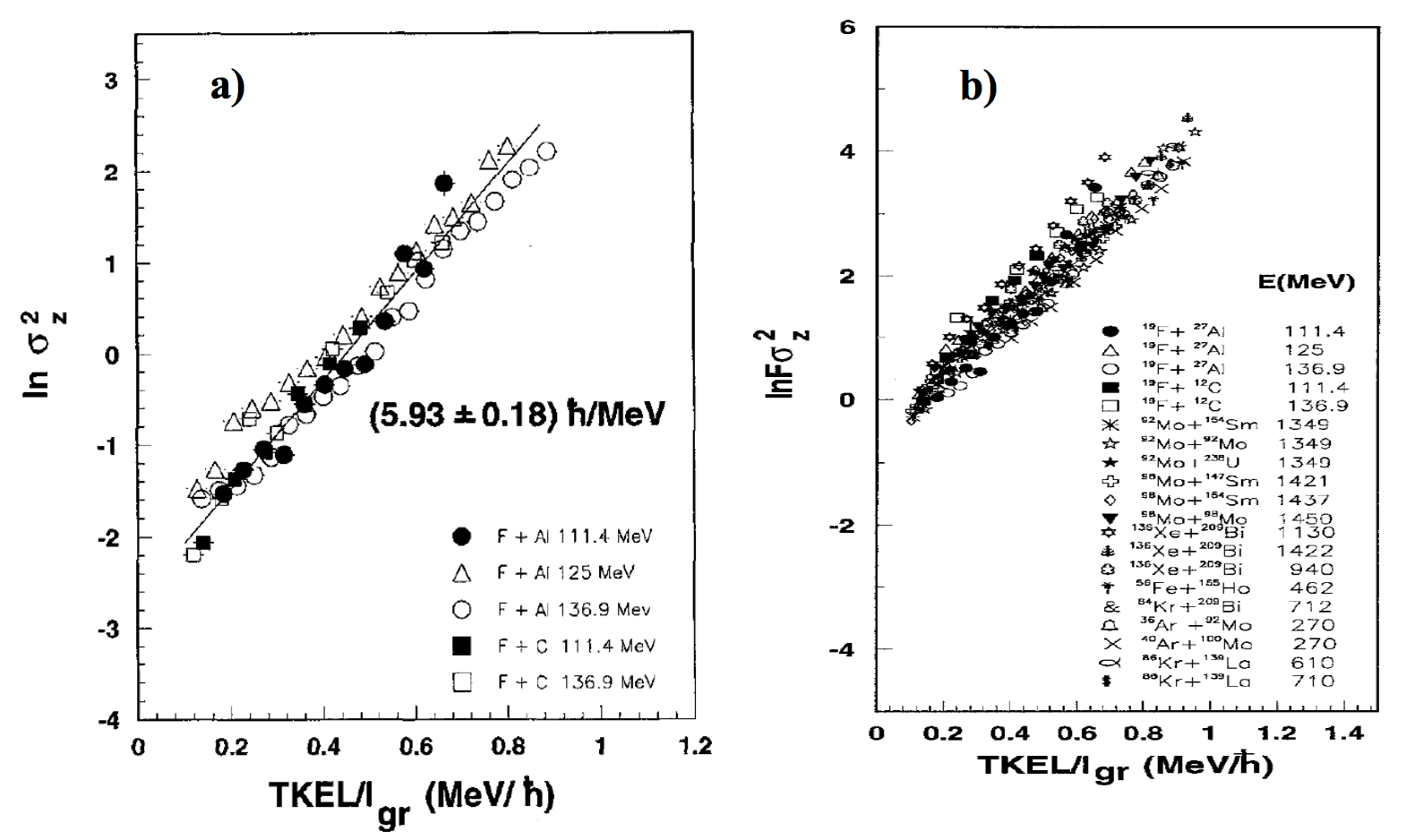}
\caption{a) ln$\sigma_Z^2$ as a function of $\frac{TKEL}{l_{gr}}$ for dissipative processes in light ion collisions at projectile energies lower then 8 A$\cdot$MeV; b) $ln(F\cdot\sigma_Z^2)$ 
as a function of $\frac{TKEL}{l_{gr}}$ for 20 different colliding systems and energies \cite{Pet1,Pet2,Pop1,Pop2,Schr,Dye,Wol1,Hoo,Birk}.}
\end{figure*}

\section{Collective expansion of compressed baryonic matter} 

 Features of collective expansion of compressed and hot baryonic matter were rather well reproduced by a hybrid model \cite{Pet5,Pet6,Pet7} combining
an analytic ideal-fluid hydrodynamic model for isentropic expansion of a spherical gas cloud in vacuum \cite{Sed,Bond} with a statistical 
disassembly model \cite{Fai} at the break-up moment. Besides a very well reproduction of the yields of different species and their average kinetic energy per nucleon in Au-Au collisions at 
250 A$\cdot$MeV projectile energy, a very interesting sequentiality in the fragment emission as a function of position within the expanding fireball was predicted. Such predictions
were confirmed by HBT studies \cite{Kot} and azimuthal distribution of the average kinetic energy for different species and of the collective expansion energy \cite{Sto,Sto1}. 
\begin{figure}[h!]
\includegraphics[width=0.95\linewidth]{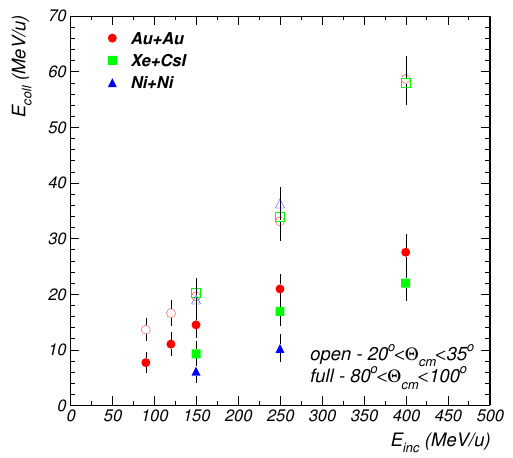}
\caption{Collective energy per nucleon as a function of projectile energy  for highly central (70 mb geometrical cross section) Au-Au, Xe-CsI and Ni-Ni collisions for $20^{\circ}\leq\Theta_{cm}\leq 35^{\circ}$-open symbols and $80^{\circ}\leq\Theta_{cm}\leq 100^{\circ}$-full symbols \cite{Sto,Pet8}}.
\end{figure}
   As far as this hybrid model is based on the similarity solutions for the equation of motion, the continuity equation and the equation of state for the isentropic expansion, for the same initial distribution of the baryonic density and temperature within the fireball as a function of self-similarity variable $x=r/R(t_b)$, the expansion velocity at the break-up moment is expected to be independent on the size of the fireball, i.e. the mass of the symmetric colliding systems, especially for central collisions. The collective energy per baryon number extracted from mean kinetic energy as a function of mass for highly central collisions for three symmetric systems Ni-Ni, Xe-Cs(I) and Au-Au between 90 and 400 A$\cdot$MeV projectile energy measured by FOPI Collaboration \cite{Sto1,Pet8} have evidenced for the same projectile energy rather different collective energy for the three systems, the values increasing from Ni-Ni to Au-Au, see Fig.3 full symbols. These experimental findings at first look seem to be in contradiction with the expectations based on the above mentioned hybrid model. As could be followed in Fig.3, these results were obtained considering the Z=1,2 and 3 species emitted at $80^{\circ}\leq\Theta_{cm}\leq 100^{\circ}$. If the analysis is done in a similar way but in $20^{\circ}\leq\Theta_{cm}\leq 35^{\circ}$ range, the results are rather different. They are  presented in Fig.3 by open symbols. For $20^{\circ}\leq\Theta_{cm}\leq 35^{\circ}$ the extracted $"$collective energy$"$ is much larger and system independent. This is the result of significant contamination from
corona processes or fluctuations in the impact parameter selection, especially for lighter systems. Therefore, although this analysis was based on highly central collision selection, i.e. 70 mb geometrical cross section, the initial configuration of the fireball is rather different in the three systems. This fact, coined as stopping, can be quantified by the ratio of the slope of ($\langle E_{kin}\rangle$) as a function of mass of the corresponding species for $80^{\circ}\leq\Theta_{cm}\leq 100^{\circ}$ relative to $20^{\circ}\leq\Theta_{cm}\leq 35^{\circ}$. An alternative receipt to quantify the stopping was based on {\it vartl}, the ratio of the variances of transverse relative to the longitudinal rapidity distributions \cite{Rei}. It was shown that the stopping is increasing significantly from Ca-Ca to Au-Au combination at a given projectile energy per baryonic number. These results clearly evidence that for a realistic extraction of collective contribution to the observed experimental trends, analysis around $\Theta_{cm}= 90^{\circ}$ is mandatory, any other polar angular range introducing a large bias in the extracted values. In the same time, it seems that similar initial configuration in terms of density and heat distribution of the fireball as a function of self-similarity variable $x=r/R(t_b)$
cannot be selected based on centrality selection due to a mass dependence of the stopping power.

\section{Geometrical scaling} 

   In a recent paper \cite{Pet3}, we made a study of the dependence
of the average transverse momentum ($\langle p_T\rangle$) on the square root of the hadron 
multiplicity over unit of rapidity and unit of  transverse overlapping area of the colliding hadrons
($\sqrt{(dN/dy)/S_{\perp}}$) for Au-Au collisions at RHIC energies and for Pb-Pb collisions at LHC energies, in order to see in which extend the expectations based on local parton-hadron duality picture (LPHD) \cite{LPHD} and  dimensionality argument 
\cite{Lap1, Lev1} are confirmed. While at RHIC energies a clear linear dependence 
of $\langle p_T\rangle$ on $\sqrt{(dN/dy)/S_{\perp}}$ was evidenced, at LHC energies a deviation from linearity towards very central collisions seems to set in, especially for protons. It was also evidenced that for an unbiased conclusion on the 
slope of $\langle p_T\rangle$ as a function of 
$\sqrt{(dN/dy)/S_{\perp}}$ dependence on collision energy and centrality, the corona contribution to the observed trends has to be considered \cite{Pet3,Pet4}. The slopes of $\langle p_T\rangle$ as a function of 
$\sqrt{(dN/dy)/S_{\perp}}$ corresponding to the core 
are systematically 
smaller relative to the values obtained based on experimental data and the difference between the values corresponding to the highest RHIC energy and the LHC energies is reduced. The slope of $\langle p_T\rangle$ as a function of the mass of pions, kaons and protons and $\langle \beta_T\rangle$, the average value of transverse expansion extracted from  
Boltzmann-Gibs Blast Wave (BGBW) fits of the
$p_T$ spectra of pions, kaons and protons,
scale as a function of $\sqrt{(dN/dy)/S_{\perp}}$ for Au-Au, starting from the lowest collision energy studied within the BES program, up to the largest collision energy in Pb-Pb collision at LHC. A similar systematics was observed in the comparison of Pb-Pb and pp colliding systems, at LHC energies, supporting the idea that at LHC energies the global evidenced trends are determined by the properties of flux tubes
of $1/\sqrt{(dN/dy)/S_{\perp}}$ size, the system size being less important.  

\begin{table}[h]  
\caption{The colliding system, the collision energy, the centrality, the average number of participant nucleons in the collision ($\langle N_{part} \rangle$), the overlapping areas corresponding to the wounded nucleons ($S_{\perp}^{geom}$)
and the hadron density 
($dN/dy$) for Xe-Xe at $\sqrt{s_{NN}}$=5.44~TeV measured by ALICE Collaboration \cite{ALICExe1, ALICExe2}.}
\label{tab:a}
\begin{tabular}{cccccccc}
\hline
\bf System & $\sqrt{s_{NN}}$ & Cen. & $<N_{part}>$ &  
$S_{\perp}^{geom}$ &  dN/dy \\ 
 &  (GeV) & ($\%$) &  & $(fm^2)$  &\\ 
\hline

     &    & 0-5& 237.5$\pm$0.0 & 121.2$\pm$0.7 & 2062.5$\pm$132.5 \\
     &    & 5-10& 207.1$\pm$0.0 & 107.7$\pm$0.7 & 1675.8$\pm$129.9 \\        
     &    &10-20& 165.0$\pm$0.1 &  89.4$\pm$0.7 & 1287.6$\pm$76.7 \\        
     &    &20-30& 118.6$\pm$0.1 &  69.7$\pm$0.7 &  878.1$\pm$55.8 \\        
Xe-Xe&5440&30-40&  82.2$\pm$0.0 &  53.7$\pm$0.7 &   581.5$\pm$42.0 \\        
     &    &40-50&  54.5$\pm$0.0 &  40.3$\pm$0.7 &  369.3$\pm$22.0 \\        
     &    &50-60&  33.7$\pm$0.0 &  28.5$\pm$0.7 &   218.9$\pm$14.0 \\        
     &    &60-70&  19.5$\pm$0.0 &  18.3$\pm$0.7 &   120.5$\pm$7.8 \\        
     &    &70-80&  10.9$\pm$0.0 &   9.9$\pm$0.6 &   42.5$\pm$2.8 \\        
\hline       
\end{tabular}        
\end{table}
\begin{figure*}[th!]
\includegraphics[width=0.95\linewidth]{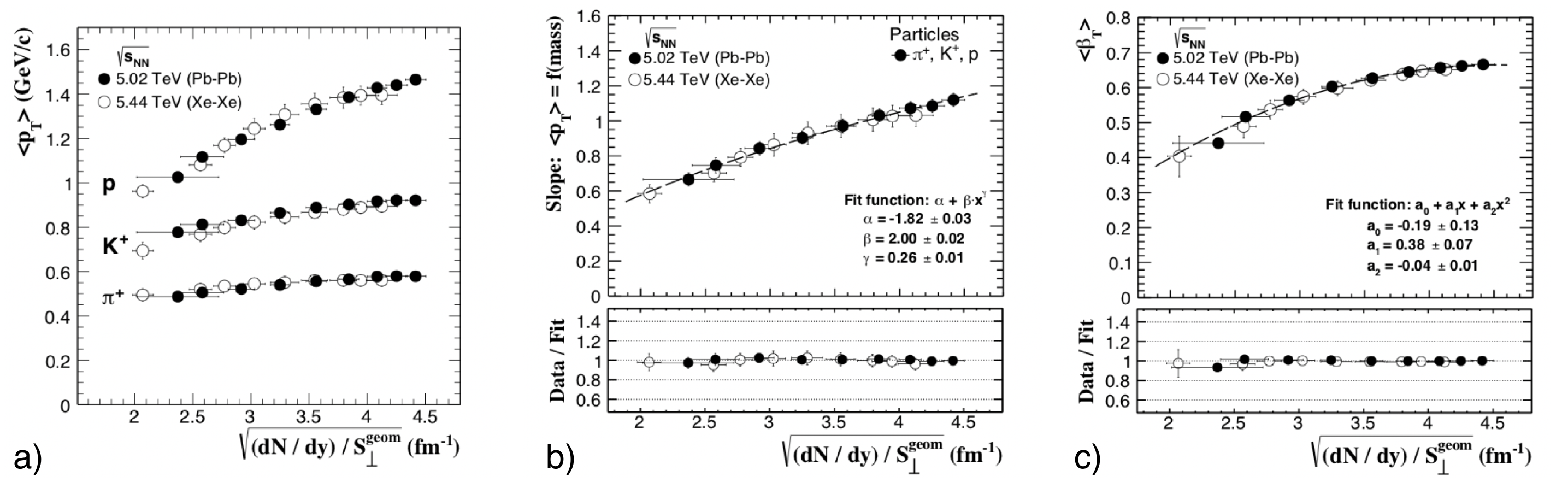}
\caption{a) $\langle p_T\rangle$; b) The slope of $\langle p_T\rangle$ as 
a function of mass of pions, kaons and protons and c) The average value of transverse expansion, $\langle \beta_T\rangle$, based on Boltzmann-Gibs Blast Wave (BGBW) simultaneous fits of 
$p_T$ spectra of pions, kaons and protons
for Xe-Xe collisions at $\sqrt{s_{NN}}$=5.44 TeV - open symbols and Pb-Pb collisions at $\sqrt{s_{NN}}$=5.02 TeV - full symbols.
}
\end{figure*}
As it is well known, recently, at LHC, Xe beam was delivered and experimental data for Xe-Xe collisions at $\sqrt{s_{NN}}$=5.44 TeV became available \cite{ALICExe1, ALICExe2}. In order to study the system size dependence of the geometrical scalings for $\langle p_T\rangle$,
the slope of $\langle p_T\rangle$ as a function of mass of pions, kaons and protons and the average value of transverse expansion, $\langle \beta_T\rangle$, based on Boltzmann-Gibs Blast Wave (BGBW) fits of the
$p_T$ spectra of pions, kaons and protons at LHC energies, one needs first to estimate 
the overlapping areas corresponding to the wounded nucleons 
($S_{\perp}^{geom}$) and the hadron density ($dN/dy$) following the recipes
explained in \cite{Pet3}. The results are presented in Table 1.

 The $\langle p_T\rangle$,
the slope of $\langle p_T\rangle$ as a function of mass of pions, kaons and protons and the average value of the transverse expansion, $\langle \beta_T\rangle$, based on Boltzmann-Gibs Blast Wave (BGBW) fits of 
$p_T$ spectra of pions, kaons and protons for Xe-Xe collisions at 
$\sqrt{s_{NN}}$=5.44 TeV and Pb-Pb collisions at $\sqrt{s_{NN}}$=5.02 TeV are presented in Fig.4 by open and full symbols, respectively.

 As could be seen from the three plots of Fig.4, all three observables for the two symmetric systems rather different in mass, at slightly different collision energies, scale very well as a function 
of charged particle density per unit of rapidity and unit of overlapping area of the colliding nuclei. The quality of the scaling of the slope of $\langle p_T\rangle$ as 
a function of mass of pions, kaons and protons and the average value of transverse expansion, $\langle \beta_T\rangle$, based on Boltzmann-Gibs Blast Wave (BGBW) fits of 
$p_T$ spectra as a function of $\sqrt{(dN/dy)/S_{\perp}}$
can be followed in the bottom plots of Fig.4b and Fig.4c, respectively, where the ratios of the data points to the fit of the experimental trends using 
the fit functions and the values of the parameters inserted in the figures. 
\begin{table}[h]  
\caption{The colliding system, the collision energy, the centrality, the average number of participant nucleons in the collision ($\langle N_{part} \rangle$), the overlapping areas corresponding to the wounded nucleons ($S_{\perp}^{geom}$), 
and the hadron density 
($dN/dy$) for Cu-Cu collision at $\sqrt{s_{NN}}$=200 GeV \cite{BRAMHS, STARCu}.}
\label{tab:b}
\begin{tabular}{cccccccc}
\hline
\bf System & $\sqrt{s_{NN}}$ & Cen. & $<N_{part}>$ &  
$S_{\perp}^{geom}$& dN/dy \\ 
 &  (GeV) & ($\%$) &  & $(fm^2)$&\\ 
 \hline
     &     &  0-10  & 99.1$\pm$0.0 & 65.9$\pm$0.6 & 328.6$\pm$13.6\\
     &     & 10-30  & 62.3$\pm$0.0 & 46.1$\pm$0.6 & 190.7$\pm$ 8.1\\
Cu-Cu& 200 & 30-50  & 30.3$\pm$0.0 & 26.1$\pm$0.6 &  85.1$\pm$ 3.5\\ 
     &     & 50-70  & 12.8$\pm$0.0 & 11.3$\pm$0.6 &  34.6$\pm$ 1.5\\ 

\hline
     
\end{tabular}        
\end{table}
\begin{figure*}[th!]
\includegraphics[width=0.95\linewidth]{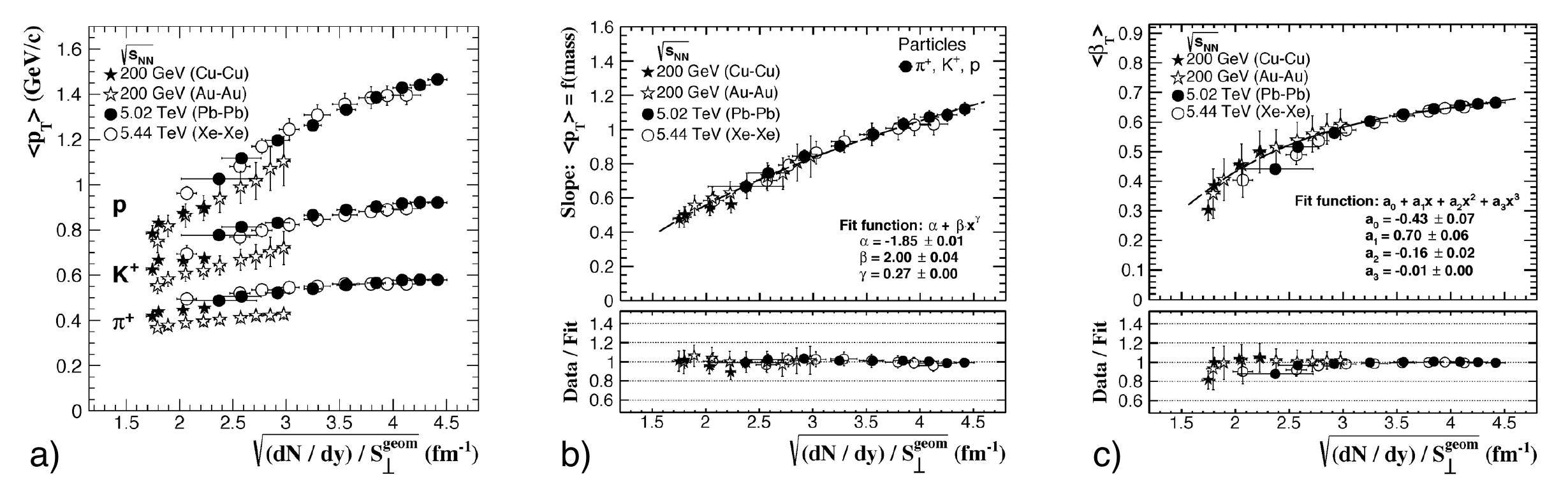}
\caption{a) $\langle p_T\rangle$; b) The slope of $\langle p_T\rangle$ as 
a function of mass of pions, kaons and protons and c) The average value of the transverse expansion, $\langle \beta_T\rangle$, based on Boltzmann-Gibs Blast Wave (BGBW) 
simultaneous fits of 
$p_T$ spectra of pions, kaons and protons
for Cu-Cu and Au-Au collisions at $\sqrt{s_{NN}}$=200  GeV - full and open stars and Xe-Xe at $\sqrt{s_{NN}}$=5.44 TeV - open bullets and Pb-Pb collisions at 
$\sqrt{s_{NN}}$=5.02 TeV - full bullets.
}
\end{figure*}
 In order to see in which extend such a scaling is confirmed for lighter
symmetric system and much lower collision energy, we used the results obtained at RHIC by 
BRAHMS and STAR Collaborations for Cu-Cu collision at $\sqrt{s_{NN}}$=200 GeV \cite{BRAMHS, STARCu}.
The average number of participant nucleons in the collision ($\langle N_{part} \rangle$), the overlapping areas corresponding to the wounded nucleons ($S_{\perp}^{geom}$) 
and the hadron density 
($dN/dy$) for Cu-Cu collision at $\sqrt{s_{NN}}$=200 GeV at measured centralities are listed in Table 2.
 Based on these values and experimental results, we obtained the results presented in Fig.5, using the same type of representations as in Fig.4, including this time also the results for Cu-Cu and Au-Au at $\sqrt{s_{NN}}$=200 GeV.

 While the $\langle p_T\rangle$ values for Cu-Cu are slightly shifted towards larger values relative to those corresponding to Au-Au at the same energy, the shift decreasing from pions to protons, the four systems show a rather good scaling going from RHIC to LHC energies in terms of the slope of $\langle p_T\rangle$ as a function of mass of pions, kaons and protons and the average value of transverse expansion as a function of $\sqrt{(dN/dy)/S_{\perp}}$.
 In the bottom plots of Fig.5b and Fig.5c are represented the ratios of the data points to the fit of the experimental trends using the functions and parameters inserted in each figure.
 These results support the conclusion of the previous paper \cite{Pet3} that the global features observed in heavy ion collisions starting from lowest collision energy at RHIC, up to the highest collision energy at LHC, are mainly dependent on the produced charged particle per unit of rapidity and unit of overlapping area.
 The size of the colliding nuclei plays a less important role. The influence of the core-corona interplay in such type of systematics as a function of mass of the colliding nuclei is under investigation. 

\section{Conclusions} 

In this lecture we presented three examples of similarities as a function of system size, evidenced or expected to be evidenced in heavy ion collisions in three very different energy ranges, from energies well below the Fermi energy, through the values where the dynamics of the compressed and hot baryonic matter plays the main role up to the LHC energies where the parton distribution of nucleons at low x and $Q^2$ starts to play the main role. At very low energies, where the mean filed of a dinuclear system and stochastic exchange of nucleons among the two partners play the main role, the variance of the exchanged nucleons scales with the ratio of the amount of energy transferred from the relative motion to the internal excitation divided by the grazing angular momenta which reflects the properties of the initial phase, i.e. collision energy and mass of the colliding system. With increasing the collision energy, the mean field phenomena becomes less important, nucleon-nucleon collisions start to be the dominant mechanism, the participant-spectator approach sets in and fireballs with high excitation energy and baryon density in the initial stage of the collision start to be formed. A hybrid model which explains rather well the main experimental results predicts for similar initial profiles in the heat energy and baryon density as a function of 
self-similarity variable $x=r/R(t_b)$ for different fireball sizes similar break-up expansion.
Deviation from such an expectation observed in the collective expansion for highly central collisions for different mass of the symmetric colliding systems shows that due to rather different stopping in light systems relative to the heavier ones, the centrality selection is not enough 
for selecting similar initial fireball profiles in terms of heat energy and baryon density as a function of self-similarity variable $x=r/R(t_b)$.
At much higher energies, starting from $\sqrt{s_{NN}}$=7.7 GeV at RHIC up to the LHC energies, the slopes of $\langle p_T\rangle$ as a function of mass for pions, kaons and protons and    
average transverse expansion, $\langle \beta_T\rangle$, extracted from simultaneous Boltzmann-Gibs Blast Wave (BGBW) fits of $p_T$ spectra of pions, kaons and protons, show a very good scaling as a function of charged particle density per unit of rapidity and unit of overlapping area. In the present contribution are presented such scalings as a function of mass of the symmetric colliding system for Cu-Cu and Au-Au collisions at $\sqrt{s_{NN}}$=200  GeV and for Xe-Xe and Pb-Pb collisions at  
$\sqrt{s_{NN}}$=5.44 and $\sqrt{s_{NN}}$=5.02, respectively.

\begin{acknowledgments}
This work was carried out under contracts sponsored by the Ministry of Research and Innovation: RONIPALICE- 04/16.03.2016, HICOR-DEFEND-F04/16.09.2016 (via Institute of Atomic Physics Coordinating Agency) and PN-18 09 01 03.  
\end{acknowledgments}
\bibliographystyle{unsrt}

\end{document}